\documentclass [11pt] {article}
\usepackage[dvips]{graphicx,color}
\usepackage{cite}
\usepackage{harvard}

\newcommand{\lchoose}[2]{{#1\choose #2}}
\newcommand{\bfc}{{\bf c}}

\newcommand{\bfx}{{\bf x}}
\newcommand{\bfy}{{\bf y}}
\newcommand{\bfone}{{\bf 1}}

\newcommand{\calk}{{\cal K}}

\newcommand{\calp}{{\cal P}}

\newcommand{\calz}{{\cal Z}}

\newcommand{\smallfrac}[2]{\frac{\mbox{\footnotesize $#1$}}{\mbox{\footnotesize $#2$}}}
\newcommand{\bfsig}{{\bf p}}
\definecolor{Red}{rgb}{1,0,0}
\definecolor{Blue}{rgb}{0,0,1}

\begin{document}
\title{
  \bf A particulate basis for a lattice-gas model of amphiphilic fluids}
\author{
  Peter J. Love\\
  {\footnotesize Centre for Computational Science,}\\
  {\footnotesize Queen Mary,
    University of London,}\\
  {\footnotesize Mile End Road, London E1 4NS, U.K.}\\
  {\footnotesize{\tt P.J.Love@qmul.ac.uk}}\\[0.3cm]
}
\date{\today}
\maketitle

\begin{abstract}
  We show that the flux-field expansion derived by~\citeasnoun{bib:micrork} for the Rothman-Keller immiscible fluid model can be derived in a simpler and more general way in terms of the completely symmetric tensor kernels introduced by those authors. Using this generalised flux-field expansion we show that the more complex amphiphilic model of~\citeasnoun{bib:bce} can also be derived from an underlying model of particle interactions. The consequences of this derivation are discussed in the context of previous equilibrium Ising-like lattice models and other non-equilibrium mesoscale models. 
\end{abstract}
\vspace{0.2truein}
\par\noindent {\bf Keywords}: Rothman-Keller model, immiscible fluids,
surfactants, lattice gases.
\section{Introduction}

Modelling and simulation of amphiphilic fluids remains a challenging area of physical and computational science. In equilibrium, ternary amphiphilic fluids self-assemble into a multitude of mesophases. The morphology of each mesophase depends on the molecular character of the amphiphile as well as on temperature, concentration, and solvent salinity. In turn, the mesophase morphology influences the nonequilibrium properties of the fluid. Wormlike micellar and sponge phases may exhibit visco-elastic rheological properties, shear induced phase transitions or demixing under flow. For a comprehensive review of amphiphilic behaviour and (mostly equilibrium) modelling techniques see~\citeasnoun{bib:gs2}.

These examples illustrate the central challenge posed by the study of amphiphilic fluids: how does one construct models which are sufficiently computationally
tractable to reach the time and length scales of interest ($\mu$m - cm) while including sufficient detail from the molecular scale (nm) to describe the behaviour of specific compounds?

The present article goes some way to addressing this issue. In particular, it shows how a previously phenomenological lattice-gas model of amphiphile dynamics can be related to an underlying model of particle interactions. In section~\ref{sec:hlga} we describe previous hydrodynamic lattice-gas automata models used to simulate binary immiscible and ternary amphiphilic behaviour. In section~\ref{sec:BCE} we describe the particular collision rules used in the Boghosian, Coveney and Emerton (1996) model. In section~\ref{sec:ce} we specify the form of the underlying model of particle interactions and obtain an expression for the change in system energy due to both collision and propagation. In section~\ref{sec:ff} we obtain the generalised form of the~\citeasnoun{bib:micrork} flux-field expansion and use it to show that the Boghosian, Coveney and Emerton (BCE) model is a truncation of this expansion at first order. 

\section{Hydrodynamic Lattice-Gas Automata}
\label{sec:hlga}

Lattice-gas automata have been used extensively for modelling
hydrodynamics since Frisch, Hasslacher, and Pomeau~\cite{bib:fchc}, and
Wolfram~\cite{bib:w4} showed that it is possible to simulate the
incompressible Navier-Stokes equations using discrete Boolean elements
on a lattice. The dynamics of all hydrodynamic lattice gases take place in two substeps: all the particles simultaneously {\it propagate} along their velocity vectors to arrive at a new lattice site (retaining their velocity vectors as they do so), and then {\it collide} with the other arriving particles.  The collisions are required to conserve mass and momentum.  Even under the constraints imposed by conservation laws, a multiplicity of collision outcomes are usually possible.
  
The simplest adaptation of the single phase lattice-gas which enables one to simulate multiphase phenomena is the introduction of an internal degree of freedom (termed `colour') for the particles. For example, if the particle colour takes values $(+1,-1)$, one has two phases, which are typically referred to as red and blue. In order to extend the range of multiphase behaviour captured by the lattice-gas, it is necessary to include colour dependent collisions. This was first done by Rothman and Keller in order to allow simulation of immiscible fluids~\cite{bib:rk}. Rothman and Keller introduced a {\it colour field}, which gives information about the composition of the neighbourhood of a site, and a colour flux. Those authors introduced a simple collision rule, choosing collision outcomes based on the colour flux and colour field which have the effect of inducing cohesion in the red and blue phases. 

The Rothman and Keller model has been used to investigate surface tension and interfacial fluctuations~\cite{bib:adler}, spinodal decomposition in both two and three dimensions~\cite{bib:rk,bib:appert}, and the effect of shear on phase separation in three dimensions~\cite{bib:olroth}.~\citeasnoun{bib:em1} generalised the Rothman-Keller model by including point surfactant particles, possessing a colour dipole. Those authors introduced three additional terms into the Rothman-Keller model. The form of these terms was based on the dipole-charge, charge-dipole and dipole-dipole interactions. The terms were then combined in a local Hamiltonian with arbitrary parameters, which were subsequently fixed by an extensive parameter search for canonical amphiphilic behaviour (micellisation). The derivation of their local amphiphilic Hamiltonian is described in~\cite{bib:bce} and its form is given below.
This model has been extensively studied in both two and three dimensions, and has been very successful in capturing a wide range of both equilibrium and non-equilibrium surfactant behaviour~\cite{bib:em1,bib:em2,bib:em3,bib:em4,bib:emshear,bib:bcp,bib:molsim,bib:LCB,bib:LCB2,bib:LCB3}. However, the model lacks any clear connection to a molecular description of surfactants, and as such the connection between different parameterisations of the model and different surfactant compounds remains unclear. Additionally, there exist many equilibrium lattice models (see~\citeasnoun{bib:gs2} for an overview) which are superficially very similar to the Boghosian, Coveney and Emerton (BCE) model, and it is unclear how the equilibrium phase diagram of the lattice gas model is related to the equilibrium phase diagram of those models.

Recently Boghosian and Coveney derived the Rothman-Keller model from an underlying model of particle interactions~\cite{bib:micrork}. In the limit where the ratio of the mean free path to the interaction range (denoted by $\epsilon$) is small, those authors showed that an arbitrary potential may be expanded in terms of fluxes and fields. The Rothman Keller model is a truncation of such an expansion at first order in $\epsilon$. This work enables one to reinterpret previous phenomenological lattice gas models in terms of an underlying potential. This has two consequences. Firstly, the interaction potential will have a specific range, which in turn associates a characteristic scale for one lattice spacing. Secondly, the parameterisation of the interaction potential enables one in principle to properly introduce molecular specificity. However, the derivation of the flux-field expansion from the underlying potential obtained for the Rothman and Keller model was of considerable complexity, making the direct extension of this work to more complex local Hamiltonians daunting. In fact, those authors noted that the simplicity of the final result obtained implied that a simpler derivation might exist.

In this paper we show that there is indeed a simpler derivation of Boghosian and Coveney's result. We write our initial potential in terms of the symmetric tensor kernels introduced by Boghosian and Coveney. By directly considering the Taylor expansion of the $mth$ rank symmetric tensor kernel, and utilising a recursion relation between the tensor kernels, we show that the flux-field expansion may be obtained much more directly. In addition, this derivation allows us to naturally generalise to the higher order tensor fluxes required for the amphiphilic lattice gas. The local amphiphilic Hamiltonian of the BCE model is obtained from this flux-field expansion to first order, in exactly the same way as the Rothman-Keller model was obtained for the binary fluid case. 

\section{Lattice-gas amphiphilic fluid dynamics}
\label{sec:BCE}

In this section we outline the original phenomenological BCE model. In single-phase lattice-gas dynamics a given incoming state $s$ has a set of possible outgoing states $\{s'\}$ which conserve mass and momentum. This set is referred to as the {\it equivalence class} of the state $s$. The hydrodynamics will be identical for any choice of $s'$ from $\{s'\}$, but in a multiphase model, the colour dynamics will be different. In the Rothman-Keller model, for each member of $\{s'\}$, a colour flux is computed, and the collision process is modified such that the notional colour work done by the flux against the field is minimised. Chan and Liang introduced a sampling procedure over the equivalence class members such that $s'$ is sampled from a probability density $\calp(s')$, which is modelled as the Gibbsian equilibrium corresponding to a Hamiltonian $H(s')$:
\begin{equation}
\calp(s')
=
\frac{1}{\calz}\exp\left[-\beta H(s')\right], \label{eq:beta_defn}
\end{equation}
where $\beta$ is an inverse temperature, $H(s')$ is the energy
associated with collision outcome $s'$, and $\calz$ is the
equivalence-class partition function. Whereas the Rothman-Keller model had a single term in the Hamiltonian, the BCE model has three additional terms, capturing the interaction of surfactant particles at a site with neighbouring colour charges, the interaction of colour charges at a site with neighbouring surfactant particles, and the interaction of surfactants with neighbouring surfactants. It is these four terms in the Hamiltonian which we systematically obtain in the remainder of the paper.

\section{Collisional Energetics}
\label{sec:ce}

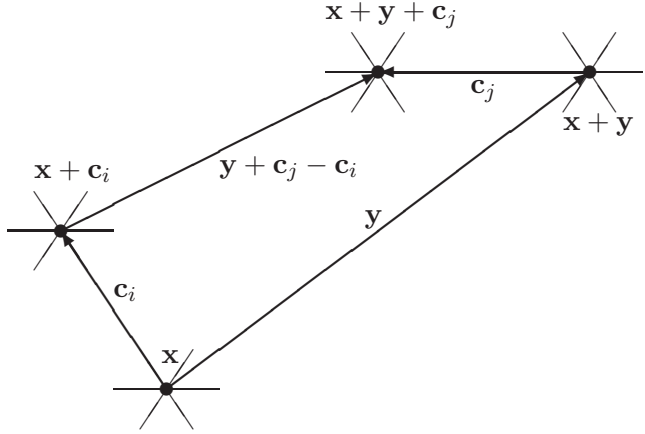
\begin{figure}
\centering
\begin{picture}(210,225)(50,25)
\put(100,60){\circle*{5}}
\put(60,120){\circle*{5}}
\put(180,180){\circle*{5}}
\put(260,180){\circle*{5}}
\put(100,60){\thicklines\vector(4,3){160}}
\put(100,60){\thicklines\vector(-2,3){40}}
\put(60,120){\thicklines\vector(2,1){120}}
\put(260,180){\thicklines\vector(-1,0){80}}
\put(100,60){\thinlines\line( 1, 0){20}}
\put(100,60){\thinlines\line( 2, 3){10}}
\put(100,60){\thinlines\line(-2, 3){10}}
\put(100,60){\thinlines\line(-1, 0){20}}
\put(100,60){\thinlines\line(-2,-3){10}}
\put(100,60){\thinlines\line( 2,-3){10}}
\put(60,120){\thinlines\line( 1, 0){20}}
\put(60,120){\thinlines\line( 2, 3){10}}
\put(60,120){\thinlines\line(-2, 3){10}}
\put(60,120){\thinlines\line(-1, 0){20}}
\put(60,120){\thinlines\line(-2,-3){10}}
\put(60,120){\thinlines\line( 2,-3){10}}
\put(180,180){\thinlines\line( 1, 0){20}}
\put(180,180){\thinlines\line( 2, 3){10}}
\put(180,180){\thinlines\line(-2, 3){10}}
\put(180,180){\thinlines\line(-1, 0){20}}
\put(180,180){\thinlines\line(-2,-3){10}}
\put(180,180){\thinlines\line( 2,-3){10}}
\put(260,180){\thinlines\line( 1, 0){20}}
\put(260,180){\thinlines\line( 2, 3){10}}
\put(260,180){\thinlines\line(-2, 3){10}}
\put(260,180){\thinlines\line(-1, 0){20}}
\put(260,180){\thinlines\line(-2,-3){10}}
\put(260,180){\thinlines\line( 2,-3){10}}
\put(98,70){$\bfx$}
\put(250,158){$\bfx+\bfy$}
\put(160,200){$\bfx+\bfy+\bfc_j$}
\put(50,140){$\bfx+\bfc_i$}
\put(175,123){$\bfy$}
\put(80,95){$\bfc_i$}
\put(215,172){$\bfc_j$}
\put(120,142){$\bfy+\bfc_j-\bfc_i$}
\end{picture}
\caption{{\bf Change in Potential Energy:} The change in the potential
  energy of interaction between two particles moving in (possibly)
  different directions at (possibly) different sites is illustrated
  here.  See Eq.~(\protect{\ref{eq:dv}}). }
\label{fig:dp}
\end{figure}

We denote the postcollision charge-like and dipolar attribute with velocity $\bfc_i$ at site $\bfx$ by $q'_i(\bfx)$ and $\bfsig'_i(\bfx)$. Upon subsequent propagation, the charges and dipoles $q'_i(\bfx)$, $\bfsig'_i(\bfx)$ will be at position $\bfx+\bfc_i$ and the charge $q'_j(\bfx+\bfy)$ and dipole $\bfsig'_j(\bfx + \bfy)$ will be at position $\bfx+\bfy+\bfc_j$.  This is
illustrated in Fig.~\ref{fig:dp}. We take as our starting point the BCE potential energy for a set of point charges and point dipoles:
\begin{eqnarray}
V 
&=&
\sum_{\bfx,\bfy}\sum_{i,j}\smallfrac{1}{2}\left(q_i(\bfx)q_j(\bfx+\bfy)\phi(|\bfy|)\right)\nonumber\\
&+& \smallfrac{1}{2}\left[q_i(\bfx) \bfsig_j(\bfx+\bfy) \cdot \bfy \phi_1(|\bfy|) - q_j(\bfx+\bfy) \bfsig_i(\bfx) \cdot \bfy \phi_1(|\bfy|)\right]\nonumber\\
&-& \smallfrac{1}{2}\left[\bfsig_j(\bfx)\bfsig_j(\bfx+\bfy):\bfy\bfy\phi_2(|\bfy|)+ \bfsig_j(\bfx)\bfsig_j(\bfx+\bfy):{\bf 1}\phi_1(|\bfy|)\right]\nonumber.\\
 \label{eq:pot}
\end{eqnarray}
Where we have defined the following functions related to
the derivatives of $\phi (y)$:
\begin{equation}
\phi_m(y)\equiv
 \left(\frac{1}{y}\frac{d}{dy}\right)^m
 \phi(y).
\end{equation}
In this case there are three contributions to the potential energy: the first term is the charge-charge interaction of the RK model, the second is the charge dipole interaction and the third term is the dipolar-dipolar interaction. We have explicitly separated the charge-dipole contribution into two terms, to make it clear that vector $i$ at site $\bfx$ may be occupied by either a charge or a dipole. The factor of $1/2$ premultiplying this term compensates for double counting.
We now introduce the completely symmetric $m$th rank kernel defined by Boghosian and Coveney:
\begin{eqnarray}
\calk_n (\bfy) & = &
 \sum_{m=\lceil n/2 \rceil}^n
 \frac{\phi_m(y)}{(n-m)!}per
 \left[
  \left(\bigotimes^{2m-n}\bfy\right)
  \otimes
  \left(\bigotimes^{n-m}\bfone\right)
 \right].
\end{eqnarray}
where ``per'' indicates a summation over all distinct permutations of indices. This enables us to write the BCE potential energy in a particularly compact form:
\begin{eqnarray}
V 
&=& 
\smallfrac{1}{2}\sum_{\bfx,\bfy}\sum_{i,j}
 q_i(\bfx)q_j(\bfx+\bfy)\calk_0 (\bfy)\nonumber\\
&+& \left[q_j(\bfx + \bfy) \bfsig_i(\bfx) - q_i(\bfx) \bfsig_j(\bfx+\bfy)\right] \cdot \calk_1(\bfy)\nonumber\\
&-& \bfsig_j(\bfx)\bfsig_j(\bfx+\bfy): \calk_2(\bfy)\nonumber\\
 \label{eq:pot2}
\end{eqnarray}
The change in the potential energy
due to {\it both} collision and propagation is then given by:
\begin{eqnarray}
\Delta V_{tot} 
&=&
\Delta V_{n} + \Delta V_{c}\nonumber\\
\Delta V_{n}
&=&
\smallfrac{1}{2}\sum_{\bfx,\bfy}\sum_{i,j}\left[q'_i(\bfx)\Delta \bfsig'_j(\bfx+\bfy)- q'_j(\bfx + \bfy)\Delta \bfsig'_i(\bfx)\right]\cdot \calk_1 (\bfy)\nonumber\\
&-&
\Delta\left[\bfsig'_j(\bfx)\bfsig'_j(\bfx+\bfy)\right]:\calk_2(\bfy)\nonumber\\
\Delta V_{c}
&=&
\smallfrac{1}{2}\sum_{\bfx,\bfy}\sum_{i,j}q'_i(\bfx)q'_j(\bfx+\bfy)\Delta \calk_0 (\bfy)\nonumber\\ 
&+& 
\left[q'_i(\bfx)\bfsig'_j(\bfx+\bfy) - q'_j(\bfx+\bfy)\bfsig'_i(\bfx)\right]\cdot \Delta \calk_1 (\bfy)\nonumber\\
&+& 
\bfsig'_j(\bfx)\bfsig'_j(\bfx+\bfy) : \Delta\calk_2(\bfy)\nonumber,\\
\label{eq:dv}
\end{eqnarray}
where we have separated the contribution to $\Delta V$ due to the movement
of the interacting particles, $\Delta V_c$, and that due to nonconservation of the dipole direction in collisions, $\Delta V_n$.

\section{The Flux-Field Decomposition}
\label{sec:ff}

Boghosian and Coveney proceeded by expanding $\Delta \calk_0(\bfy)$
in a Taylor series in the ratio of the characteristic lattice
spacing $c$ to the characteristic interaction range $y$. We wish to expand $\Delta \calk_0(\bfy)$,
$\Delta \calk_1(\bfy)$ and $\Delta \calk_2(\bfy)$ in a similar way. These first and second rank tensors are functions of the vector $\bfy$, not only of its magnitude. The Taylor expansion of the $m$th rank kernel is given by:
\begin{equation}\label{eq:taylordef}
\calk_m(\bfy + \bfc) = \sum_{n=1}^\infty \frac{1}{n!} (\bfc \cdot \nabla)^n \calk_m(\bfy).
\end{equation}
This may be written as a sum over $n$th rank inner products of $n$th rank outer products:
\begin{equation}\label{eq:taylordef2}
\calk_m(\bfy + \bfc) = \sum_{n=0}^\infty \frac{1}{n!} \biggl( \bigotimes^n \nabla \biggr) \calk_m(\bfy) \bigodot^n \bigotimes^n \bfc,
\end{equation}
enabling one to use the property of the $n$th tensor derivative of the $m$th rank kernel (which is proved in Appendix~\ref{app:proof}):
\begin{equation}\label{eqn:recursion}
\biggl (\bigotimes^n \nabla \biggr) \calk_m(\bfy) = \calk_{m+n}(\bfy).
\end{equation}
Letting $\bfc = \bfc_j - \bfc_i$ in equation~(\ref{eq:taylordef}):
\begin{eqnarray}
\calk_m(\bfy + \bfc_j - \bfc_i) 
 &=&
\sum_{n=1}^\infty \frac{1}{n!} \left[(\bfc_j - \bfc_i)\cdot \nabla\right]^n \calk_m(\bfy)
\end{eqnarray}
and binomially expanding $\left[(\bfc_j - \bfc_i)\cdot \nabla\right]^n$:
\begin{equation}
\left[(\bfc_j - \bfc_i)\cdot \nabla\right]^n = \sum_{k=0}^n \lchoose{n}{k} (-1)^k (\bfc_j \cdot \nabla)^{n-k}(\bfc_i \cdot \nabla)^{k}\calk_m(\bfy),
\end{equation}
writing $(\bfc_j \cdot \nabla)^{n-k}$ and $(\bfc_i \cdot \nabla)^{k}$ in terms of $n$th rank inner products of $n$th rank outer products as in equation~(\ref{eq:taylordef2}) gives:
\begin{eqnarray}\label{eq:kexp}
\calk_m(\bfy + \bfc_j - \bfc_i) 
 &=&
\sum_{n=0}^\infty \frac{1}{n!}\sum_{k=0}^n \lchoose{n}{k} (-1)^k \left( \bigotimes^k \bfc_i \right) \bigodot^k \calk_{m+n}(\bfy) \bigodot^{n-k} \left(\bigotimes^{n-k} \bfc_j\right).
\end{eqnarray}
Using~(\ref{eq:kexp}) to substitute for $\Delta \calk_m(\bfy)$ in $\Delta V_c$ give the charge-charge interaction as:
\begin{eqnarray}
\Delta V_{cc} 
 &=&
\smallfrac{1}{2}\sum_{\bfx}\sum_{n=1}^\infty \frac{1}{n!}\sum_{k=0}^n {\cal J}^q_k(\bfx) \bigodot^k \lchoose{n}{k} (-1)^k \sum_{\bfy} \calk_{n}(\bfy) \bigodot^{n-k} {\cal J}^q_{n-k}(\bfx+\bfy),\nonumber\\
\nonumber\\
\end{eqnarray}
and the charge-dipole interaction as:
\begin{eqnarray}
\Delta V_{cd} 
 &=&
\smallfrac{1}{2}\sum_{\bfx}\sum_{n=1}^\infty \frac{1}{n!}\sum_{k=0}^n{\cal J}^q_k(\bfx)\bigodot^k \lchoose{n}{k} (-1)^k \sum_{\bfy} \calk_{n+1}(\bfy) \bigodot^{n-k+1} {\cal J}^p_{n-k+1}(\bfx + \bfy)\nonumber\\
 &-&
\smallfrac{1}{2}\sum_{\bfx}\sum_{n=1}^\infty \frac{1}{n!}\sum_{k=0}^n{\cal J}^p_{k+1}(\bfx)\bigodot^{k+1} \lchoose{n}{k} (-1)^k \sum_{\bfy} \calk_{n+1}(\bfy) \bigodot^{n-k} {\cal J}^q_{n-k}(\bfx+\bfy)\nonumber\\
\nonumber\\
\end{eqnarray}
and the dipole-dipole interaction as:
\begin{eqnarray}
\Delta V_{dd} 
 &=&
\smallfrac{1}{2}\sum_{\bfx}\sum_{n=1}^\infty \frac{1}{n!}\sum_{k=0}^n {\cal J}^p_{k+1}\bigodot^{k+1} \lchoose{n}{k} (-1)^k \sum_{\bfy} \calk_{n+2}(\bfy) \bigodot^{n-k+1}{\cal J}^p_{n-k+1} ,\nonumber\\
\nonumber\\
\end{eqnarray}
where we have defined:
\begin{eqnarray}
{\cal J}^p_{k+1}(\bfx)
&=&
\sum_{i}\bfsig'_i(\bfx)\bigotimes^{k}\bfc_i\nonumber\\
{\cal J}^q_k(\bfx)
&=&
\sum_{i}q'_i(\bfx)\bigotimes^k\bfc_i\nonumber\\
\end{eqnarray}
Because these equations are summed over the whole lattice we have a freedom of choice in labelling the particles. This means that the above expressions are invariant under the interchange:
\begin{eqnarray}
 i &\leftarrow& j\nonumber\\
 j &\leftarrow& i\nonumber\\
 \bfx &\leftarrow&\bfx+\bfy\nonumber\\
 \bfy &\leftarrow&-\bfy
 \label{eq:symsa}
\end{eqnarray}
In the first and third terms this means that the $kth$ term and the $n-kth$ term are equal. For the charge dipole terms the $kth$ term in the $q_i \bfsig_j$ series is equal to the $n-kth$ term in the $q_j \bfsig_i$ series. We can therefore drop half of the sum over $k$. To first order in epsilon the change in potential energy due to the motion of the particles is therefore:
\begin{eqnarray}\label{eqn:firstorder}
\Delta V_c^{(1)} 
&=&
-{\cal J}^q_1(\bfx) \bigodot^1 \sum_{\bfy} \calk_{1}(\bfy){\cal J}^q_{0}(\bfx+\bfy)\nonumber\\
 &-&
{\cal J}^q_1(\bfx)\bigodot^1 \sum_{\bfy} \calk_{2}(\bfy) \bigodot^{1} {\cal J}^p_{1}(\bfx + \bfy)\nonumber\\
 &+&
{\cal J}^p_{2}(\bfx)\bigodot^{2}\sum_{\bfy} \calk_{2}(\bfy) {\cal J}^q_{0}(\bfx+\bfy)\nonumber\\
 &-&
{\cal J}^p_{2}\bigodot^{2}\sum_{\bfy} \calk_{3}(\bfy) \bigodot^{1}{\cal J}^p_{1}.
\end{eqnarray}
Introducing the definitions of colour flux:
\begin{equation}
{\bf J} = {\cal J}^q_1,
\end{equation}
colour field:
\begin{equation}
{\bf E} = -\sum_{\bfy} \calk_{1}(\bfy){\cal J}^q_{0}(\bfx+\bfy), 
\end{equation}
dipolar field:
\begin{equation}
{\bf P} = -\sum_{\bfy} \calk_{2}(\bfy)\cdot{\cal J}^p_{1}(\bfx+\bfy), 
\end{equation}
dipolar flux tensor:
\begin{equation}
{\cal J} = {\cal J}^p_2 = \sum_{i}\bfsig'_i(\bfx)\bfc_i,
\end{equation}
colour field gradient tensor:
\begin{equation}
{\cal E} = \sum_{\bfy} \calk_{2}(\bfy){\cal J}^q_{0}(\bfx+\bfy),
\end{equation}
and dipolar field gradient tensor:
\begin{equation}
{\cal P} = -\sum_{\bfy} \calk_{3}(\bfy)\cdot{\cal J}^q_{1}(\bfx+\bfy).
\end{equation}
from the BCE model. Substituting these definitions into the first order expansion~(\ref{eqn:firstorder}) yields:
\begin{equation}
\Delta V_c^{(1)} = {\bf J}\cdot({\bf E} + {\bf P}) + {\cal J}:({\cal E}+{\cal P})
\end{equation}
which is identically the BCE local Hamiltonian without the terms due to collisional non-conservation of dipolar moment. 

It only remains to evaluate $\Delta V_n$. Firstly we retain only those parts of $\Delta V_n$ which distinguish between outgoing states, writing:
\begin{equation}
\Delta V_n  = \sum_{\bfx,\bfy}\sum_{i,j}  q'_j(\bfx + \bfy)\bfsig'_i(\bfx)\cdot \calk_1 (\bfy) + 
\smallfrac{1}{2}\left[\bfsig'_j(\bfx)\bfsig'_j(\bfx+\bfy)\right]:\calk_2(\bfy)\
\end{equation}
Using the definitions (18-23) and defining the total outgoing dipole vector:
\begin{equation}
\bfsig'(\bfx) = \sum_{i}\bfsig'_i(\bfx)
\end{equation}
we obtain:
\begin{equation}\label{collchange}
\Delta V_n' = \smallfrac{1}{2}\sum_{\bfx} \bfsig'(\bfx) \cdot \left( 2{\bf E} + {\bf P}\right)
\end{equation}
This has exactly the form used in the BCE model. The factor of two above, which was not included in subsequent utilisations of the model, arises because they derived the various terms in a local Hamiltonian piecewise, without taking account of double counting considerations. These terms in the local Hamiltonian were then combined with arbitrary coefficients. 

\section{Discussion and Conclusions}

Having obtained the BCE model Hamiltonian from an underlying potential, it is now possible to obtain parameterisations of this model for a given form of interaction potential. Equally, it is possible to attempt the inverse problem and use the much studied two- and three-dimensional parameterisations of the model to obtain a form for the interaction potential. Any computational implementation of the model will utilise particular stencils to evaluate the tensor fields and gradients over a finite range. These stencils, together with the form of the Hamiltonian defined here and a given model parameterisation will yield a set of differential equations and boundary conditions for $\phi(y)$. Once obtained, this form of $\phi(y)$ could be utilised in other mesoscale models. Most notably, the amphiphilic Malevanets-Kapral described in~\cite{bib:anatolytern1} is very similar to the BCE model. This raises the interesting possibility of studying the same underlying model of particle interactions in both lattice and off-lattice models. However, previous studies of the BCE used nearest neighbour stencils. The interaction range is therefore comparable to the mean-free path (epsilon is close to one) and so additional studies of the BCE model with longer-range stencils may be necessary to make this comparison.

Additionally, many equilibrium lattice models of amphiphilic systems exist, whose phase diagrams have been obtained by Monte-Carlo simulation. One in particular, that of Matsen and Sullivan utilises the same form for the potential of a set of interacting charges and dipoles as that used here~\cite{bib:matsen1,bib:matsen2}. A given set of parameters for this equilibrium model can therefore be directly translated into a set of parameters for our non-equilibrium model. This enables the equilibrium phase diagram of a given set of parameters to be calculated by Monte-Carlo simulation, or equivalently the dynamic behaviour calculated for a set of parameters for which the equilibrium phase diagram is already known.

By giving previously phenomenological models an underlying theoretical foundation, the conceptual problems inherent in the lattice-gas approach, and in mesoscale modelling in general, are thrown into sharp relief. Although we have referred to a `local Hamiltonian', in fact the lattice gas algorithm does not pay any attention to energy conservation. One could consider following the same approach as that used for energy-conserving DPD, and attach an internal degree of freedom to the lattice gas particles~\cite{bib:consdpd}. However, such an internal energy changes the nature of the particles from microscopic to thermodynamic entities. Their dynamics should then be required not only to satisfy energy conservation, but also guarantee positive entropy production. 

The lattice-gas models studied in~\cite{bib:em1,bib:em2,bib:em3,bib:em4,bib:emshear,bib:bcp,bib:molsim,bib:LCB,bib:LCB2,bib:LCB3} have elements of both interpretations. On the one hand lattice gas surfactant particles are regarded as being genuinely molecular in character, whereas in the collision step outgoing states are sampled over Boltzmann weights constructed from the local Hamiltonian. This sampling procedure effectively puts each lattice site in contact with a heat bath, thereby introducing a thermodynamic aspect to the simulation. This thermodynamics is flawed in three ways. Firstly, as was shown by Chan and Liang, there is a net heat flow between the heat bath and the simulation, despite the fact that they are at the same notional temperature~\cite{bib:chli}. Secondly, the total energy of the heat bath plus system is not conserved, and thirdly the system does not possess a H-function. Put more simply, the thermodynamic aspects of these lattice-gas simulations violate the zeroth, first and second laws of thermodynamics.

Thermodynamically consistent mesoscale models remain elusive. At present only the GENERIC formulation of DPD can claim complete thermodynamic consistency, at the expense of abandoning any molecular level information~\cite{bib:serpep}. Such a completely top-down approach can only be useful for materials for which a widely accepted macroscopic description exists. Such materials include liquid-gas mixtures or nematic liquids, but no commonly agreed upon continuum description exists for many materials, including amphiphiles. 

The thermodynamic difficulties of the lattice-gas algorithm described above arise partly from the mixture of microcanonical and canonical elements. In the context of equilibrium Monte-Carlo simulation an approach by Creutz shows how to perform equilibrium Monte-Carlo simulations in which microcanonical and canonical elements are incorporated in a systematic way~\cite{bib:creutz}. It is possible that similar approaches to the non-equilibrium simulations considered here could also yield useful modifications.

\section*{Acknowledgements}

The author would like to thank Peter Coveney, Bruce Boghosian and Jonathan Chin for productive discussions and CINECA and the EU human mobility program for a MINOS grant during which part of this work was undertaken. Figure 1 is reproduced with permission from~\citeasnoun{bib:micrork}.

\newpage
\appendix

\section{Proof of recursion relation eqn~(\ref{eqn:recursion}).}\label{app:proof}

We now prove the recursion relation~(\ref{eqn:recursion}) by induction. We start by noting that~(\ref{eqn:recursion}) is trivially true for $n=1, m=0$ and therefore the result $\nabla \calk_n(\bfy) = \calk_{n+1}(\bfy)$ completes our proof of~(\ref{eqn:recursion}). Writing out the left hand side of this expression and using $\nabla \phi_m(y) = \bfy \phi_{m+1}(y)$ gives:
\begin{eqnarray}
\nabla \calk_n(\bfy)
&=& 
\sum_{m=\lceil n/2 \rceil}^n
 \frac{\phi_{m+1}(y)}{(n-m)!}\bfy per\left[2m-n,n-m \right]\nonumber\\
&+&
\sum_{m=\lceil n/2 \rceil}^n
 \frac{\phi_m(y)}{(n-m)!}\nabla per\left[2m-n,n-m \right],
\end{eqnarray}
where we have introduced the further shorthand:
\begin{equation}
per \left[ \left(\bigotimes^{2m-n}\bfy\right) \otimes \left(\bigotimes^{n-m}\bfone\right) \right] = per \left[2m-n,n-m \right].
\end{equation}
By substituting $m' = m-1$ in the first term here and considering series for odd and even $n$ separately, we obtain:
\begin{eqnarray}
\nabla \calk_n(\bfy)
&=&  \phantom{\sum_{m=\lceil (n+1)/2  \rceil}^{n}}\frac{\phi_{n/2}(y)}{(n/2)!} \nabla per \left[0,m \right]\nonumber\\
&+&  \sum_{m=\lceil n/2 + 1 \rceil}^{n}
 \frac{\phi_{m}(y)}{(n+1-m)!}\biggl(\bfy per\left[2m-2-n,n+1-m \right]\nonumber\\
&+& (n+1-m)\nabla per\left[2m-n,n-m \right]\biggr)\nonumber\\
&+& 
\phantom{\sum_{m=\lceil (n+1)/2  \rceil}^{n}}\phi_n+1(y) \bfy per\left[n,0 \right].
\end{eqnarray}
for even $n$ and:
\begin{eqnarray}
\nabla \calk_n(\bfy)
&=& 
\sum_{m=\lceil n/2 + 1 \rceil}^{n}
 \frac{\phi_{m}(y)}{(n+1-m)!}\biggl(\bfy per\left[2m-2-n,n+1-m \right]\nonumber\\
&+& (n+1-m)\nabla per\left[2m-n,n-m \right]\biggr)\nonumber\\
&+&
\phantom{\sum_{m=\lceil (n+1)/2  \rceil}^{n}}\phi_n+1(y) \bfy per\left[n,0 \right].
\end{eqnarray}
for odd $n$. Using $\bfy per\left[n,0 \right] = per\left[n+1,0 \right]$ and $\nabla per \left[0,m \right] = 0$ enables us to reunite these two series, giving:
\begin{eqnarray}\label{nearlythere}
\nabla \calk_n(\bfy)
&=& 
\sum_{m=\lceil (n+1)/2 \rceil}^n \frac{\phi_{m}(y)}{(n+1-m)!} \bfy per\left[2m+2-n,n+1-m \right] \nonumber\\
&+&
\sum_{m=\lceil (n+1)/2 \rceil}^n \frac{\phi_{m}(y)}{(n+1-m)!}(n+1-m) \nabla per\left[2m-n,n-m \right]\nonumber\\
&+&
\phantom{\sum_{m=\lceil (n+1)/2 \rceil}^n} \phi_{n+1}(y) per\left[n+1,0 \right].
\end{eqnarray}

We can simplify this expression if we consider the types of additional permutations of indices in $per \left[2m-n-1,n+1-m \right]$ as compared with those in 
$per \left[2m-n,n-m \right]$. Clearly, the $(n+1)th$ rank tensor has one additional index, which can be attached either to a $\bfy$ or to a Kr\"{o}necker delta. If the new index is attached to a $\bfy$ there is only one additional permutation, which is taken care of by the first term in~(\ref{nearlythere}). The second term will result in $2m-n$ terms on differentiation, the new index being combined in each successive term in a Kr\"{o}necker delta with each of the $2m-n$ indices labelling the $\bfy$'s in the original expression. These terms capture all the unique ways of assigning the new index to a Kr\"{o}necker delta. 

The prefactor $(n+1-m)$ is necessary as the whole expression is divided by  $(n+1-m)!$, which arises as this factor is exactly the number of ways of permuting $n+1-m$ Kr\"{o}necker deltas, this corresponds to $(n+1-m)!$ equivalent permutations of the indices. These permutations are not generated by the outer derivative, but this factor captures their effect on the resulting tensor, enabling us to write:
\begin{eqnarray}
\biggl(\bfy per\left[2m-2+n,n+1-m \right]+ \nabla per\left[2m-n,n-m \right]\biggr)=per \left[2m-n-1,n+1-m \right]
\end{eqnarray}
Substitution of this result into~(\ref{nearlythere}) immediately yields the result $\nabla \calk_n(\bfy) = \calk_{n+1}(\bfy)$ and so completes our proof.

\end{document}